\documentclass[twocolumn,showpacs,preprintnumbers,amsmath,amssymb,nofootinbib,pra]{revtex4}
\usepackage{graphicx}
\usepackage{dcolumn}
\usepackage{bm}
\usepackage{amsmath}
\usepackage{ulem}

\newcommand{\bi}{\begin{itemize}}
\newcommand{\ei}{\end{itemize}}
\newcommand{\bt}{\begin{theorem}}
\newcommand{\et}{\end{theorem}}
\newcommand{\bp}{\begin{proof}}
\newcommand{\ep}{\end{proof}}
\newcommand{\be}{\begin{equation}}
\newcommand{\ee}{\end{equation}}
\newcommand{\ben}{\begin{enumerate}}
\newcommand{\een}{\end{enumerate}}

\newcommand{\C}{\mathbb C}

\newcommand{\R}{\mathbb R}

\newcommand{\Rscr}{\mathcal R}

\newcommand{\hf}{\frac{1}{2}}

\newcommand{\e}{\varepsilon}

\newcommand{\m}{\mu}

\renewcommand{\o}{\omega}

\renewcommand{\t}{\tau}

\newcommand{\z}{\zeta}

\newcommand{\D}{\Delta}
\renewcommand{\S}{\Sigma}
\renewcommand{\a}{\alpha}

\newcommand{\g}{\gamma}

\renewcommand{\and}{\text{~~ and ~~}}
\renewcommand{\part}{\partial}
\newcommand{\ra}{\rightarrow}

\newcommand{\ie}{{i\over \e}}

\newcommand{\sech}{{\rm sech~}}

\newcommand{\ipt}{\frac{i\pi}{2}}

\renewcommand{\D}{\Delta}

\newcommand{\rv}{\mathbf{r}}

\begin{document}

\title{Semiclassical dynamics of quasi-one-dimensional, attractive Bose-Einstein condensates}
\author{Alexander Tovbis}
\email{atovbis@pegasus.cc.ucf.edu}
\affiliation{Department of Mathematics, University of Central Florida, Orlando, Florida 32816}
\author{M. A. Hoefer}
\affiliation{Department of Mathematics, North Carolina State
  University, Raleigh, NC 27695}

\begin{abstract} 
  The strongly interacting regime for attractive Bose-Einstein
  condensates (BECs) tightly confined in an extended cylindrical trap
  is studied.  For appropriately prepared, non-collapsing BECs,
  the ensuing dynamics are found to be governed by the one-dimensional
  focusing Nonlinear Schr\"{o}dinger equation (NLS) in the
  semiclassical (small dispersion) regime.  In spite of the
  modulational instability of this regime,  some mathematically
  rigorous results on the strong asymptotics of the semiclassical
  limiting solutions  were obtained recently. Using these results,
  ``implosion-like'' and ``explosion-like'' events are
  predicted whereby an initial hump focuses into a sharp spike which
  then expands into rapid oscillations.  Seemingly related behavior
  has been observed in three-dimensional experiments and models, where
  a BEC with a sufficient number of atoms undergoes collapse.
    The dynamical regimes studied here,
  however, are not predicted to undergo collapse.  Instead, distinct,
  ordered structures, appearing after the ``implosion'', yield
  interesting new observables that may be experimentally
  accessible.


\end{abstract}

\pacs{
  03.75.Kk, 	
  05.45.Yv      
}

\maketitle

\section{Introduction}
\label{sec:introduction}

Bose-Einstein condensates (BECs) with attractive interactions between
atoms have been found to exhibit collapse \cite{Gerton} resulting in
violent, three-dimensional (3D) explosive dynamics
\cite{Donley,Cornish} and the propagation of quasi-one-dimensional
(1D), stable bright solitary waves \cite{Strecker,Khaykovich}.  The
implosion and subsequent explosion of a 3D BEC, a ``Bosenova'', has
been explained theoretically as a blow-up singularity of the
Gross-Pitaevskii (GP) equation when the condensate has a sufficient
number of atoms \cite{Ruprecht,Tsurumi,Gammal}.  A quasi-1D soliton
train was formed by exploiting this instability \cite{Strecker}.
Other pattern forming instabilities in BECs include modulational and
transverse instability \cite{Kevrekidis04} which result in the
formation of coherent, localized nonlinear structures such as solitons
and vortices \cite{Anderson}.  In this work, we explore a new and
completely different mechanism leading to violent dynamics and the
formation of quasi-periodic nonlinear matter wave trains, the
semiclassical (zero dispersion limit) regime where a quasi-1D
attractive BEC can ``implode'' and ``explode'' yet does not undergo 3D
collapse.  This dynamical regime can be accessed with a cigar shaped
BEC with negative s-wave scattering length of sufficiently small
magnitude.  Recent experiments suggest that this may be possible
\cite{Pollack}.


Exact, analytical results for the \textit{strong asymptotics} of the
small dispersion limit of the 1D focusing (attractive)
Nonlinear Schr\"{o}dinger equation (NLS) are used to describe the
onset of a gradient catastrophe or sharp focusing of the condensate
density (implosion) followed by breaking (explosion) \cite{TVZ1,KMM}.
We emphasize, however, that these dynamics do not correspond to
  collapse of the 3D BEC wavefunction.  Rather, the resulting
dynamics reveal two counterpropagating radiative waves, the space
between them filled by rapidly oscillating quasi-periodic (2-phase)
nonlinear matter waves. Further dynamics are determined by the
discrete spectrum (solitons) of the associated Zakharov-Shabat
eigenvalue problem for the 1D focusing NLS. In the absence of the
discrete spectrum, the pure radiation case, the amplitudes of the
(2-phase) nonlinear matter waves are decaying exponentially in time
\cite{TVZ2}. In the presence of the discrete spectrum, (the number of
points is inversely proportional to the semiclassical parameter) more
complicated localized structures ($2n$-phase nonlinear waves) can
appear within the oscillatory region with $n > 1$ growing in time.
The presence of an initial, sufficiently large inward BEC superfluid
velocity (phase gradient) can completely remove the localized coherent
structures in the oscillatory region \cite{TVZ1}.

In this work, we apply the aforementioned rigorous results to an
attractive quasi-1D BEC in a cigar shaped trap.  The small dispersion
regime can be accessed by condensing a sufficiently large number of
repulsive atoms in a properly sized cigar shaped trap.  As has been
done in the past \cite{Donley,Cornish,Strecker,Khaykovich,Pollack}, a
Feshbach resonance can then be applied to tune the sign of the
nonlinear term, resulting in an attractive mean field interaction.
While avoiding collapse of the entire condensate, the 1D implosion and
explosion events are shown to be experimentally accessible for
sufficiently tight radial confinement.  Various observables resulting
from the dynamics of the localized, prepared condensate are elucidated
including the point of gradient catastrophe (breaking point), curves
in the space-time plane, separating different asymptotic regimes
(breaking curves), see Fig.\ref{Cai}, and the asymptotic structure of
the BEC density.

The outline of the paper is as follows.  First, we introduce the
appropriate parameter regime for the application of the semiclassical
NLS equation to a BEC.  Following this, we review recent rigorous
results on the focusing NLS equation with small dispersion
and discuss their application to single hump initial conditions.

\section{1D BEC as a semiclassical limit of the focusing NLS}
\label{sec:1d-bec-as}

The temporal evolution of a BEC is governed by the GP equation for the
condensate wave function $\Psi(\mathbf{r},t)$, given by
\cite{Pethick01,Pitaevskii03}
\begin{equation}
\label{GPEfull}
i\hbar \frac{\partial}{\partial
  t}\Psi=\left[-\frac{\hbar^2}{2m}\nabla^2 + V(\rv,t)
  +g(t) \vert\Psi\vert^2\right]\Psi\,, 
\end{equation}
where: $m$ is the single atom mass, $V = V_\perp(y,z) +
V_\parallel(x,t)$ is an external trapping potential with radial
$V_\perp$ and axial $V_\parallel$ terms, $g(t) = 4\pi\hbar^2 a_s(t)/m$
is the nonlinear coefficient arising due to two-particle interactions
and is characterized by the scattering length $a_s$.  We will assume
that
\begin{equation}
  \label{eq:26}
  V_\parallel(x,t) = \left \{
    \begin{array}{cc}
      V_\parallel(x) & t < 0 \\
      0 & t \ge 0
    \end{array} \right ., \quad a_s(t) = \left \{
    \begin{array}{cc}
      a_s^{(r)} > 0 & t < 0 \\
      a_s^{(a)} < 0 & t \ge 0
    \end{array} \right . .
\end{equation}
For $t < 0$, the BEC is repulsive (positive scattering length) and
confined in all three spatial dimensions.  At $t=0$, the axial
confinement is turned off and the scattering length is rapidly
switched to a negative value, e.g.~by a Feshbach resonance
\cite{Pethick01}, so that the condensate becomes attractive.

We assume that the BEC has been prepared (for $t < 0$) in the ground
state of a strongly anisotropic trap
\begin{equation}
  \label{eq:25}
  V_\perp(y,z) = \frac{1}{2} m \omega_r^2(y^2 + z^2) ,
\end{equation}
where $\omega_r$ is the harmonic trap frequency with radial
localization width ${a_0 = \sqrt{\hbar/m
    \omega_r}}$
The axial portion of the
potential $V_\parallel(x)$, for $t < 0$, confines the BEC's axial
extent to a width $\Delta \gg a_0$.

Equation (\ref{GPEfull}) conserves the particle number
\begin{equation}
  \label{eq:3}
  \int_{\R^3} |\Psi(\rv,t) |^2 d \rv = N .
\end{equation}
A sufficient condition to \textit{avoid} collapse in a harmonic
potential is to take $N$ sufficiently small $N < N_\textrm{cr}$
\cite{Carles02}.  In two spatial dimensions with a harmonic trap, it
has been shown that $N_\textrm{cr}$ is related to the loss of
stability of the nonlinear ground state \cite{Tsurumi,Berge00}.  Using
stability theory for nonlinear ground states \cite{Rose88}, one can
compute an estimate of $N_\textrm{cr}$.  This result has been assumed
to hold for 3D BECs as well, leading to numerical calculations of
$N_\textrm{cr}$ for several trap configurations
\cite{Berge00,Gammal,Parker07}.  For the 3D harmonic potential
$V(x,y,z) = V_\perp(y,z)$ that we are considering at $t\geq 0$, a
numerical calculation in \cite{Parker07} determined
\begin{equation}
  \label{eq:27}
  N_\textrm{cr} \approx 0.67 \frac{a_0}{|a_s^{(a)}|} .
\end{equation}

We now consider the one-dimensional reduction of eq.~(\ref{GPEfull})
for $t > 0$.  This simplification requires that the characteristic
energies of the radial excitations are much greater than the energies
associated with the axial and nonlinear excitations
\cite{Abdul,Kevrekidis}.  This regime leads to the following
requirement 
\begin{equation}
  \label{eq:15}
  C\frac{N |a_s^{(a)}|}{\Delta} \ll 1 ,
\end{equation}
where the $O(1)$ factor $C$ can be estimated through the initial
density of the BEC (see further discussion in the Appendix).  If
(\ref{eq:15}) is satisfied, then $\Psi$ in eq.~(\ref{GPEfull}) for $t
> 0$ can be approximated by \cite{Abdul,Kevrekidis}
\begin{equation}
  \label{eq:29}
  \Psi(\xi,y,z,t) \approx \sqrt{\frac{N}{\Delta}} \phi(y,z) q(\xi/\Delta,t/k) e^{-i
    \omega_r t}
\end{equation}
where $\phi(y,z) = \exp[-(y^2+z^2)/(2 a_0^2)]/(\sqrt{\pi}a_0)$ is the
2D, linear ground state for the transverse harmonic potential.  Here,
the axial variable $x$ has been replaced by $\xi$ for convenience.
The remaining axial and temporal dependence is embodied in the
function $q(x,\tau)$ which satisfies the NLS equation
\begin{equation}
  \label{FNLS}
  i\varepsilon q_\t + \frac{\varepsilon^2}{2} q_{xx}+\vert
  q\vert^2 q=0~,
\end{equation}
where, with a slight abuse of notation, we re-introduce $x=\xi/\Delta$
which is now non-dimensional and set $\tau=t/k$.  The parameters in
eq.~(\ref{FNLS}) are
\begin{equation}
  \label{eq:31}
  \e = \frac{a_0}{\left (2 N |a_\textrm{s}^{(a)}| \Delta \right )^{1/2}}, \quad   k =
  \D\left (\frac{m\D}{2|a_s^{(a)}|\o_r N\hbar} \right )^{1/2} . 
\end{equation}
We are interested in the semiclassical (small dispersion) regime where
the semiclassical parameter $\e$ satisfies
\begin{equation}
  \label{eq:30}
 0 < \e \ll 1 .
\end{equation}
Conservation of particle number in (\ref{eq:3}) combined with
eq.~(\ref{eq:29}) gives
\begin{equation}
  \label{eq:28}
  \int_\R |q(x,\t)|^2 dx = 1 .
\end{equation}

In summary, we have derived the NLS equation (\ref{FNLS}) in the small
dispersion regime with the assumptions of $N < N_\textrm{cr}$ (\ref{eq:27})
and the inequalities (\ref{eq:15}) and (\ref{eq:30}).  Experimentally,
all of the parameters $a_s^{(a)}$, $N$, $\Delta$, and $a_0$ can be
varied so that we expect that these three inequalities can be
satisfied.  By direct calculation, we find that all three inequalities
(\ref{eq:15}), (\ref{eq:30}), and $N < N_\textrm{cr}$ can be satisfied
by choosing, for example,
\begin{equation}
  \label{eq:48}
  N = \frac{a_0}{2 |a_s^{(a)}|}, \quad \sqrt{\frac{a_0}{\Delta}} \ll
  1, \quad \frac{C a_0}{2 \Delta} \ll 1 . 
\end{equation}
The factor $C$ here (and in \eqref{eq:15}) is chosen so that $\int_\R
|q(x,\t)|^4dx < C$.  The inequalities in (\ref{eq:48}) are just the
requirement of cigar shaped initial data with tight radial
confinement.  Using a Feshbach resonance, one can, in principle, tune
the scattering length $a_s^{(a)}$ to an arbitrary value so that
eq.~(\ref{eq:48}) implies that any number of atoms is possible. In a
recent experiment, the scattering length for $^7$Li has been precisely
tuned over seven orders of magnitude \cite{Pollack}.  This
demonstrates that the integrable, 1D NLS equation in the small
dispersion limit may be a valid model for BEC experiments.

To illustrate the above discussion, consider the following data for
the bright soliton experiments in \cite{Strecker} with $^7$Li.  The
parameters involved are $m \approx 10^{-26}$ kg, $a_s^{(a)} \approx
-1.6 \cdot 10^{-10}$ m, $a_0 \approx 1.6 \cdot 10^{-6}$ m, and $N
\approx 3 \cdot 10^5$.  With these parameters, $N_\textrm{cr} \approx
6700$ and the system is predicted to undergo collapse, as observed in
the experiment.  Nevertheless, assuming an initial axial width of
$\Delta \approx 200 \, \mu$m, which is the approximate experimental
value, we calculate
\begin{equation}
  \label{eq:1}
  \e\approx 0.011, \quad k \approx 14 \, \textrm{ms}, \quad
  N |a_s^{(a)}|/\Delta \approx 0.24. 
\end{equation}
Equation (\ref{eq:1}) ensures the validity of the semiclassical
regime for at least  short times, 
whereas the quasi-1D assumption of
eq.~(\ref{eq:1}) is, perhaps, on the borderline of applicability.

\section{Summary of predictions from rigorous asymptotic analysis}
\label{sec:summ-pred-from}

\begin{figure}
  \includegraphics[width = 0.49\textwidth]{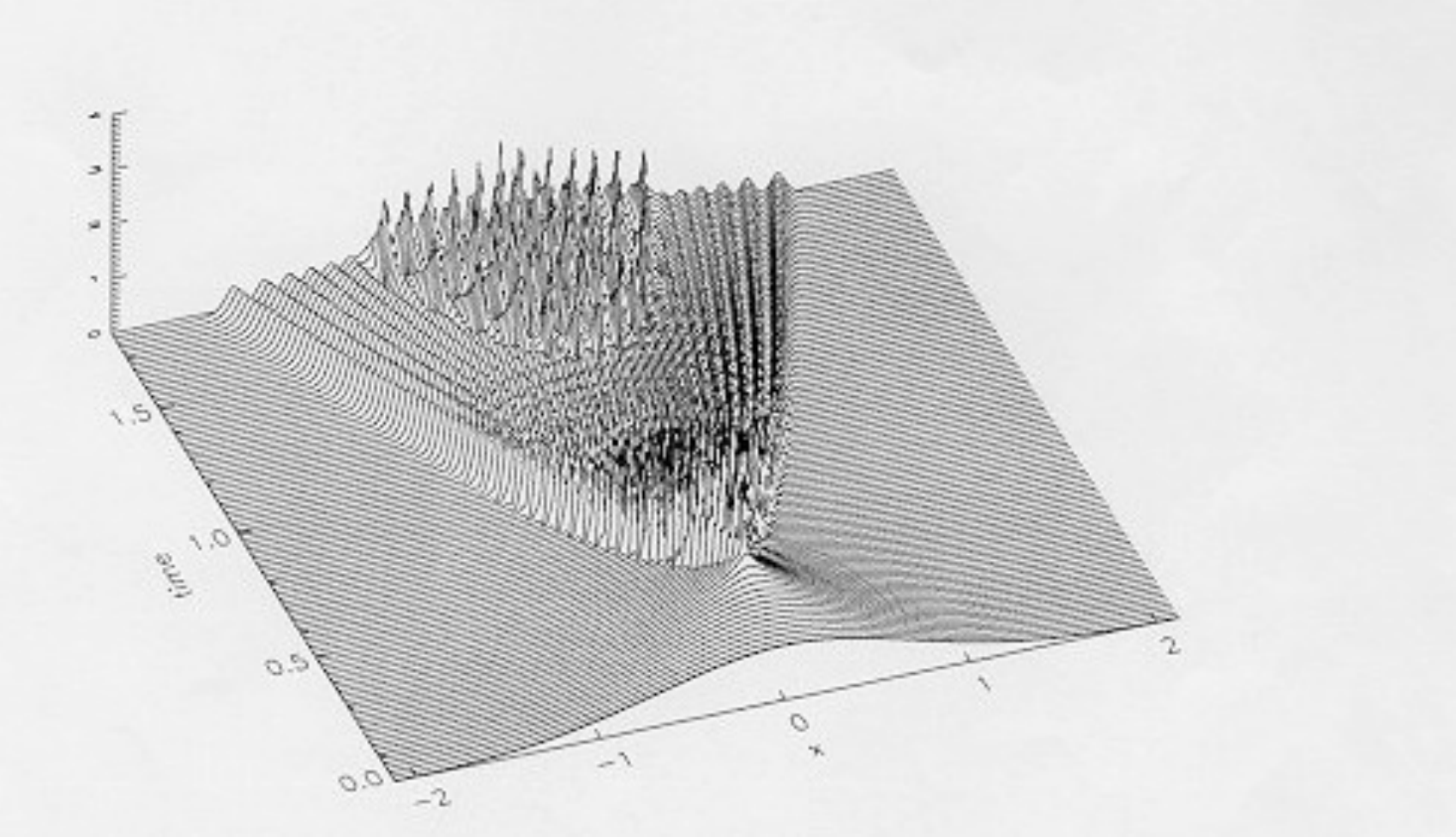}
  \caption{Absolute value $|q(x,\t,\varepsilon)|$ of a solution
    $q(x,\t,\varepsilon)$ to the focusing NLS (\ref{FNLS}) versus
    $x,\t$ coordinates.  Here the initial data
    \eqref{IDe} is given by $A(x)=e^{-{x^2}}$, $S'(x)=-\tanh x$, and
    $\e=0.02$. Reprinted from Handbook of Dynamical Systems, Vol 2, 
D. Cai, D. W. McLaughlin, and T. T. R. McLaughlin, The nonlinear Schrodinger equation as both a 
PDE and a dynamical system, pp 599-675, Copyright (2002) with permission from Elsevier.}
\label{Cai}\end{figure}

Here we bring together and summarize some of the main results from the
semiclassical, rigorous asymptotic analysis undertaken in
\cite{TVZ1,TV0,TVZ2,TVZ3,TV1,TV2,TV3}.  Details can be found in
Sections \ref{sec:semicl-limit-solut} and \ref{sec:calc-semicl-nls}.
Figure \ref{Cai} from \cite{CMM} depicts implosion and explosion
dynamics generated by the single-hump initial condition $q(x,0) =
e^{-x^2} \exp (-\frac{i }{ \e} \ln\cosh x )$ for the NLS equation
(\ref{FNLS}).  Very similar dynamics were rigorously derived in
\cite{TVZ1} for the one parameter family of initial data
\begin{equation}
  \label{eq:2}
  q(x,0,\e) = \sech x\,  e^{-\frac{i \mu}{2 \e} \ln\cosh x },
\end{equation}
where $\mu\geq 0$ provides a measure of the phase gradient $(\arg
q)_x$ or inward superfluid velocity in the context of BEC
\cite{Pethick01}.  The key features of this study, illustrated by
Figure \ref{Cai}, can be extended to generic decaying analytical
one-hump initial data.  here are some highlights:
\begin{itemize}
\item The physical $x,\t$ plane is subdivided into regions where the
  solution is asymptotically (as $\e\ra 0$) described by modulated
  $2n$-phase nonlinear waves (the $n=0$ or plane wave approximation
  corresponds to the smooth region in Fig.~\ref{Cai}, $n=1$ to the
  next oscillatory region, etc.).
\item Phase transitions between regions of different behavior are
  separated by \textit{breaking curves} in the $x$, $\tau$ plane that
  do not depend on $\e$.  Equations for the breaking curves are given
  by \eqref{br1}. A detailed description of the transitional behavior
  at the breaking curve can be found in \cite{BT1};
\item The tip of the breaking curve $(x_0,\tau_0)$ is the point of
  gradient catastrophe for the plane wave approximation.  Behind this
  tip, the solution suddenly bursts into rapid amplitude oscillations
  or density spikes.  Each spike within the vicinity of $(x_0,\tau_0)$
  has the height $3|q(x_0,\t_0)|$ and the shape of the rational
  breather solution to the NLS, while the locations of the spikes
  correspond to the poles of the special tritronqu\'ee solution to the
  first Painlev\'{e} equation, see \cite{BT2}.  For the family
  \eqref{eq:2}, the exact location of the point of gradient
  catastrophe is $(x_0,\t_0)=(0,\frac{1}{\m+2})$ with $| q(x_0,\t_0) |
  = \sqrt{\mu + 2}$ so that the height of the spikes are
  $3\sqrt{\m+2}$; the slope of the breaking curve is
  $\frac{\cot\frac{\pi}{5}}{\sqrt{\m+2}}$.
\item The asymptotic solution for $q$ in the plane wave approximation
  region is completely characterized by the implicit formulas in
  eq.~(\ref{ab2}) when $\mu = 2$ and eq.~(\ref{ab0}) when $\mu = 0$.
  In the general case, it is determined by eq.~(\ref{modeqg0}).
\item When the initial data \eqref{eq:2} has $\mu \le 2$, there exist
  $\mathcal{O}(1/\e)$ points of discrete spectrum (solitons) centered
  at $x = 0$.  Their interaction leads to the secondary break starting
  at about $(x,\tau) \approx (0,1.5)$ in Fig.~\ref{Cai}, followed by
  the region of 4-phase wave approximation.
\item When the initial data \eqref{eq:2} has $\mu > 2$, there are no
  solitons.  Therefore, no secondary breaking region exists and the
  solution contains only two regions of distinct behavior: smooth and
  oscillatory.  The smooth region ($n=0$) decays exponentially fast in
  time to zero whereas the background of the oscillatory region decays
  as $\mathcal{O}(\tau^{-1/2})$ (however, the amplitude of the spikes
  decays exponentially to the background with $\t$).  A large inward,
  focusing momentum prevents the formation of higher breaking regions.
\end{itemize}

\section{Semiclassical limit solutions to the focusing NLS}
\label{sec:semicl-limit-solut}

In this subsection we  discuss some mathematical
background and recent analytical results related to the semiclassical
regime of the focusing NLS.  The focusing Nonlinear Schr\"odinger
equation \eqref{FNLS} is a universal, basic model for self-focusing
and self-modulation in that it describes the evolution of the envelope
of modulated waves in generic weakly nonlinear, dispersive systems.
It is also one of the most celebrated nonlinear integrable equations
that was first integrated by Zakharov and Shabat \cite{ZS}, who used
the inverse scattering procedure to describe general decaying
solutions ($\lim_{|x|\to \infty}q(x,0)=0$) in terms of radiation and
solitons.  Central to their discovery was the so-called Lax pair which
effectively linearizes the NLS equation.  The first equation in the
Lax pair for the NLS is called the Zakharov-Shabat (ZS) system. It is
used to define the correspondence between the initial data and the
scattering data. Since the scattering data undergoes an explicit and
very simple time evolution, the inverse scattering transform (IST), 
which maps the scattering data back to physical space,
is used to obtain the evolution of given initial data at any
time $\t>0$.

In the semiclassical limit ($\varepsilon \to 0$), the focusing NLS
\eqref{FNLS} exhibits {\it modulationally unstable} behavior (see
Fig. \ref{Cai}), as was first shown in \cite{FL}.  This is in drastic
contrast to the case of the defocusing (repulsive) NLS equation
\cite{CMM,KKU} in which the semiclassical theory shows regions of
modulated periodic or quasi-periodic oscillation. These two very
different types of behavior can be explained through the modulation
equations, which are elliptic in the focusing case and hyperbolic in
the defocusing case. The corresponding initial value problems are,
therefore, ill-posed and well-posed respectively.  As a result, a
plane wave with amplitude modulated by $A(x)$ and phase modulated by
$S(x)$, taken as initial data \be \label{IDe}
q(x,0,\varepsilon)=A(x)e^{iS(x)/\varepsilon} \ee
for the focusing NLS \eqref{FNLS}, is expected to break immediately
into disordered oscillations of both the amplitude and the phase.
However, in the case of \textit{analytic} initial data, the NLS
evolution displays some orderly structure instead of the disorder
suggested by the modulational instability, see
\cite{CMM,MK,CT}. Throughout this work, we will use the abbreviation
NLS to mean ``focusing Nonlinear Schr\"odinger equation".

Figure \ref{Cai} from \cite{CMM} depicts the time evolution of a
typical Gaussian-shaped, symmetric, analytic initial data \eqref{IDe}.
It identifies regions where different types of behavior (modulated
$2n$-phase nonlinear waves) of the solution $q(x,\t,\varepsilon)$
appear. In particular, consecutive regions with $n=0,2$ and,
presumably, $n=4$ are depicted in Fig.~\ref{Cai}.  These regions are
separated by curves in the $x,\t$ plane that are called breaking
curves or nonlinear caustics.  The location of a breaking curve is
defined by $A(x)$ and $S(x)$ from \eqref{IDe}; it does not depend on
$\e$.  Within the $2n$-phase wave approximation region, the
\textit{strong asymptotics} of $q(x,\t,\varepsilon)$ can be expressed
in terms of Riemann Theta-functions (see, for example, \cite{TVZ1}),
that are defined on the genus $2n$ hyperelliptic Riemann surface
$\Rscr(x,\t)$.  Therefore, the $2n$-phase wave approximation region is
referred to as the genus $2n$ region.  Because of the symmetry of the
ZS system, $\Rscr(x,\t)$ is Schwarz-symmetrical.  The surface
$\Rscr(x,\t)$ and, more precisely, its $2n+1$ complex branch points
(because of the symmetry, we consider only branch points in the upper
half-plane), {\it do not depend} on the semiclassical parameter
$\e$. They can be viewed as slowly varying functions of the space-time
variables that describe the wave's parameters, i.e., the {\it
  macroscopic structure} of the solution in the vicinity of a given
point $x,\t$.

Equations that define the branch points of $\Rscr(x,\t)$ are known as
{\it modulation} or Whitham equations. In the case $n=0$ (genus zero
case), $\Rscr(x,\t)$ has only two branch points: $\a=\a(x,\t)$ and its
complex conjugate $\bar\a$.  In this case, the Riemann Theta-function
expression for $q(x,\t,\varepsilon)$ is replaced by
\be\label{qg0} q(x,\t,\varepsilon)=A(x,\t)e^{iS(x,\t)/\varepsilon}+O(\e),
\ee where \be\label{alph} \a(x,\t)=-\hf S_x(x,\t)+i A(x,\t), 
\ee 
and $A(x,0)=A(x)$, $S(x,0)=S(x)$.  The genus zero region is the first
region adjacent to the $\t=0$ axis where the solution \eqref{qg0} has
the form of a high frequency $\mathcal{O}(1/\e)$ modulated wave with
slowly varying amplitude $A(x,\t)$ and phase $S(x,\t)$.

For the initial data \eqref{eq:2}, the corresponding scattering data
for \eqref{FNLS} was calculated explicitly \cite{TV0} (For general
initial data of the type \eqref{IDe} see
Sec.~\ref{sec:calc-semicl-nls}). Using this data, the modulated
amplitude and phase of \eqref{qg0} and \eqref{alph} can be obtained
from the system of transcendental equations for
$\a(x,\t)=a(x,\t)+ib(x,\t)$ :
\begin{equation}
\label{transc}
\begin{cases}
&\sqrt{(a-T)^2+b^2}+\sqrt{(a+T)^2+b^2}=\m+2\t b^2, \\
&\left[a-T+\sqrt{(a-T)^2+b^2}\right]\left[a+T+\sqrt{(a+T)^2+b^2}\right]\\
&=b^2 e^{2(x+2\t a)}~,
\end{cases}
\end{equation}
where $T=\sqrt{\frac{\m^2}{4}-1}$.  It is interesting to mention here
that, in general, the phase gradient $S'(x)$ has a significant impact
on the asymptotic behavior of the evolving solution. For example
\cite{TV0}, in the case $0 \le \m<2$, the corresponding ZS eigenvalue
problem has $O\left(1/\e\right)$ points in the discrete spectrum
(solitons) located on the vertical segment $[-T,T]$. These solitons
are localized at $x=0$, do not move, and their interaction can produce
quite complicated coherent structures in the process of evolution.  In
Fig.~\ref{Cai}, these structures are seen after the second break
(around $x=0$ and $\t>1.5$), which, presumably, corresponds to the
genus 4 region.  It is expected that the solution undergoes more phase
transitions (breaks) for larger values of time $\t$ that are not
visible in Fig.~\ref{Cai}.  On the contrary, in the case $\m>2$, the
ZS eigenvalue problem does not have any discrete spectrum (solitons).
The evolution of such initial data produces only two asymptotic
regimes (of genera zero and two), similar to the first two asymptotic
regions in Fig.~\ref{Cai} (the second asymptotic region extends to
$t=\infty$, see \cite{TVZ1} for the proof). The breaking curve
separating these two regions has linear (slanted) asymptotes as
$\t\ra\infty$.  In the limit $\t\ra \infty$, the solution in the genus
zero region approaches zero exponentially fast, whereas in the genus
two region (inside the wedge) it decays as $O(\t^{-\hf})$ \cite{TVZ1}.
The high frequency amplitude oscillations in this region decay
exponentially in $\t$, and the solution has the profile of a parabola
with a maximum at $x=0$ and zero values along the breaking curve.
Qualitatively, these results mean that a large focusing momentum
(directed towards the origin), generated by the phase gradient
$S'(x)=-\m \tanh x$ with $\m>2 $, prevents the formation of the higher
genera regions.
 
In the borderline case $\m=2$, equations \eqref{transc} have a
particularly simple solution.  Introducing the implicit time
$u=u(x,\t)$ at each point $x\in\R$ by $\t=(u-x)[\sinh
2u-(u-x)]/(4\sinh^2 u)$, one can obtain an explicit solution of
eq.~(\ref{transc}) (see \cite{TVZ1})
\begin{equation}
  \label{ab2} 
  a=\frac{2\sinh^2 u}{\sinh
    2u-(u-x)}~,~~~~~b=\frac{2\sinh u}{\sinh 2u-(u-x)}~
\end{equation}
for $A(x,\t)=b$ and $S'(x,\t)=-2a$ that are valid throughout the genus
zero region. Similar expressions with the implicit time $u=u(x,\t)$
given by $\t=\hf\sqrt{(u-x)[\sinh 2u-(u-x)]}\coth u$ and
\begin{equation}
  \label{ab0}
  a^2=\frac{(u-x)\tanh^2 u}{\sinh 2u-(u-x)}~,~~~~~b^2=\frac{2\tanh
    u}{\sinh 2u-(u-x)}~ 
\end{equation}
are valid in the case $\m=0$.

Notice that the amplitude $A(x,t)$ of the solution in Fig.~\ref{Cai}
at first contracts (focuses) towards the point of maximum amplitude, $x=0$,
and then suddenly bursts into {rapid (order $1/\e$) and} violent
oscillations, transitioning to the genus two regime.
This is typical behavior \cite{TVZ3} for an analytic one-hump initial
condition provided that $S'(x)$ does not decrease too fast.  The very
first point of this transition, which is the tip-point of the first
breaking curve (see Fig. \ref{Cai}), is called a point of {\it
  gradient catastrophe} \cite{DGC}.  At the point $(x_0,\t_0)$ of the
gradient catastrophe, the semiclassical solution \eqref{qg0} of
\eqref{FNLS} loses its smoothness \cite{TV3}, i.e.,
$\a_x(x_0,\t_0)=\infty$ (either $A_x(x,\t_0)$ or $S_{xx}(x,\t_0)$ or
both become infinite). The recent results of \cite{BT2} show that each
of the spikes seen in Fig.~\ref{Cai} immediately after the moment of
gradient catastrophe represents a rational breather (see
\cite{ratbreath}) solution to the NLS.  The height of each spike is
exactly three times the value of the amplitude at the time of gradient
catastrophe, i.e., $3A(x_0,\t_0)+O(\e^{1/5})$ and the location of the
spikes are determined by the poles of the \textit{tritronqu\'ee}
solution of the first Painlev\'{e} equation.

\begin{figure*}
\includegraphics[width = 2\columnwidth]{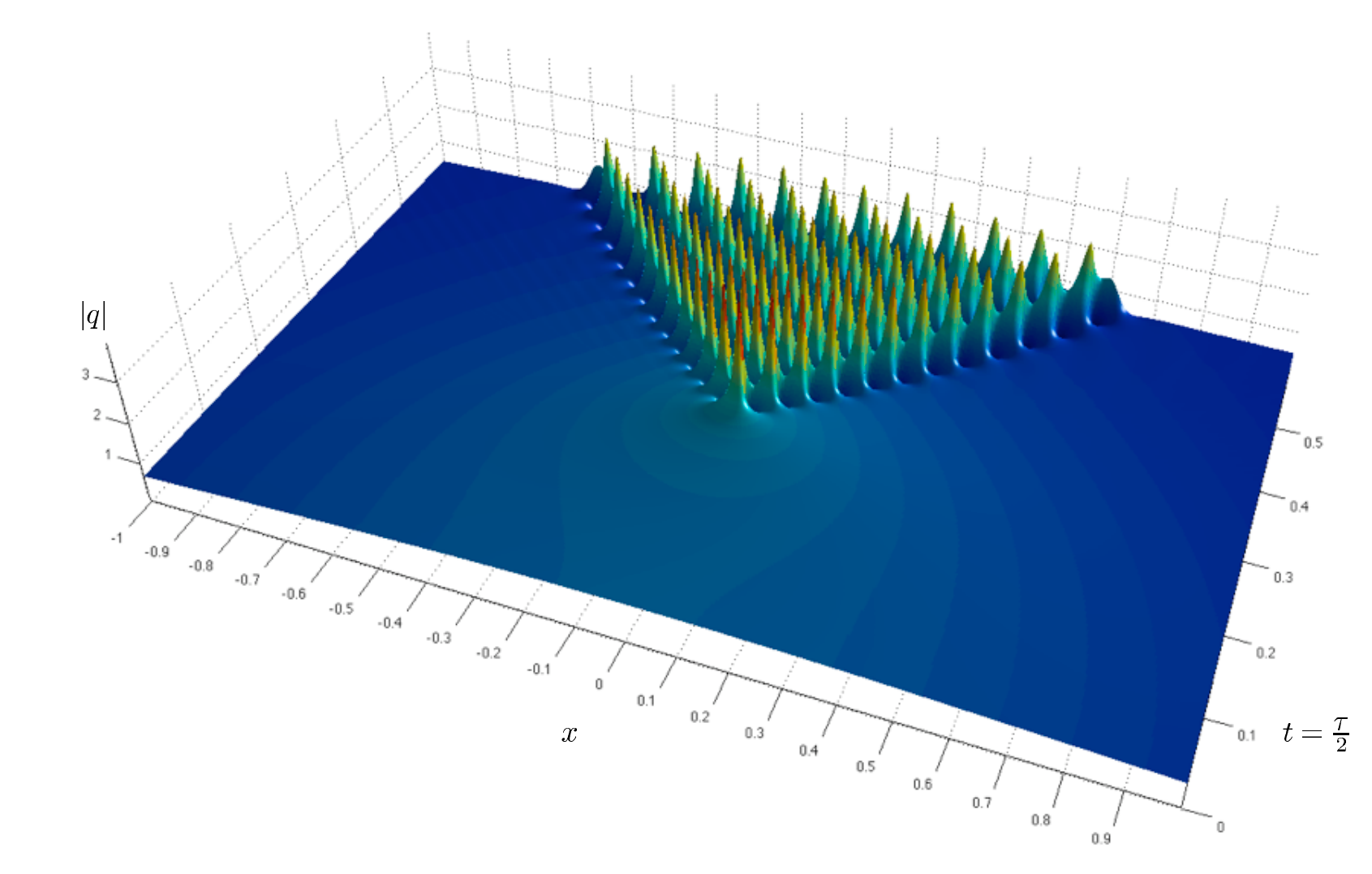}
\caption{Numerical simulation of the solution of \eqref{FNLS} with the
  initial data \eqref{eq:2}, where $\m=0$, $\e=1/33$. The time scale
  $t$ used here and our time $\t$ are related through $\t=2t$. This
  simulation confirm the values $\t_0=0.5$ and $b_0=\sqrt{2}$. It also
  shows that each spike near the point of gradient catastrophe has the
  height of $3b_0$ (with the accuracy $(1/33)^{1/5}\approx 0.496$) and
  the shape of a rational breather.}
  \label{numerics}
\end{figure*}

Due to the symmetry of the initial data \eqref{eq:2}, the gradient catastrophe occurs
at $x_0=0$. In the cases $\m=2$ and $\m=0$, the time of the gradient catastrophe 
$\t_0$ can be calculated as $\t_0=\lim_{u\ra 0}\t(u,0)$, where the expressions
for $\t=\t(u,x)$ are given above formulae \eqref{ab2} and \eqref{ab0} respectively.
This yields $\t_0=1/4$ for $\mu=2$ and $\t_0=1/2$ for $\mu=0$. The value of the
amplitude $b_0=A(x_0,\t_0)$ at the point of gradient catastrophe can be calculated as
$b_0=\lim_{u\ra 0}b(u,0)$, where $b=b(u,x)$ are given in \eqref{ab2} and in \eqref{ab0}.
Thus, $b_0=2$ for $\m=2$ and $b_0=\sqrt{2}$ for $\m=0$. In the case $\m=0$,
numerical solution of \eqref{FNLS} with $\e=1/33$ is shown on Fig. \ref{numerics}.

\begin{figure}
  \includegraphics[width = 0.49\textwidth]{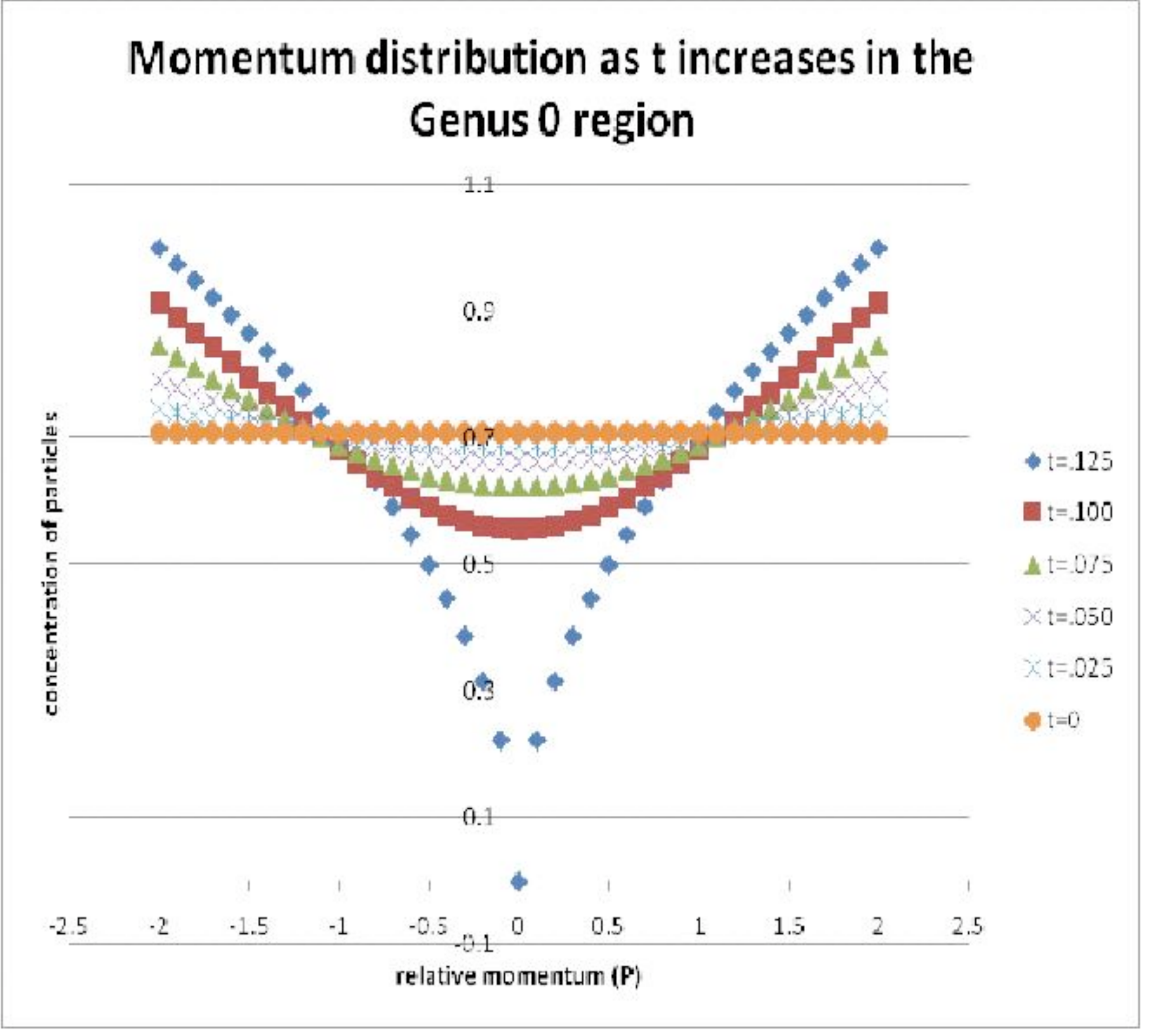}
  \caption{Evolution of the Fourier transform  of the initial data
    \eqref{IDe} with $A(x)=\sech x$ and $S'(x)=-2\tanh x$ in the limit
    $\e\ra 0$ from $\t=0$ to the time of gradient catastrophe
    $\t=0.25$; time $t$ shown on the figure and $\t$ are related by
    $\t=2t$. The vertical axis shows the values of $\frac{1}{\sqrt{\e}}|\hat q|$, the 
horizontal axis shows the  relative momentum $P$, which  is defined as $P=\e p$, where $p$ is the standard momentum.
The data shown here is explicitly calculated from \eqref{Fourier} through the standard stationary point method.}
  \label{Benfig}
\end{figure}

One possible mechanism for the loss of one-dimensionality
  described by our 1D analysis is the generation of large axial
  momentum.  This may correspond to the breaking down of the energetic
  assumption of weak axial kinetic energy relative to the radial
  harmonic oscillator energy (see eqs.~(\ref{eq:15}) and
  (\ref{eq:7})).  To illustrate this point, we consider the evolution
of the Fourier transform 
\be\label{Fourier}
\hat q(p,\t,\e)=\frac{1}{\sqrt{2\pi}}\int_\R A(x,\t)e^{\ie[S(x,\t)+\e px]}dx
\ee
 of the solution $q(x,\t,\e)$ to \eqref{FNLS}  in the genus zero region
(where $q(x,\t,\e)$ has the form \eqref{qg0}).
If we define the relative momentum $P$  through the standard momentum
$p$ as $P=\e p$, the standard stationary phase method can be applied to 
\eqref{Fourier}. Time evolution of $\frac{1}{\sqrt{\e}}\left|\hat q(P/\e,\t,\e)\right|$ that corresponds 
to the  initial data \eqref{IDe} with $A(x)=\sech x$ and $S'(x)=-2\tanh x$ 
is shown on Fig. \ref{Benfig} (in this case  $A(x,\t)$ and  $S(x,\t)$ are given
by \eqref{alph} and \eqref{ab2}). In fact, direct calculations show that in the leading $\e$-order
$\frac{1}{\sqrt{\e}}\left|\hat q(P/\e,0,\e)\right|=\frac{1}{\sqrt{2}}$ and 
$\frac{1}{\sqrt{\e}}\left|\hat q(P/\e,1/4,\e)\right|= \sqrt{|P|/2}$ when $|P|\leq 2$ and 
$\frac{1}{\sqrt{\e}}\hat q(P/\e,\t,\e)=0$ when $|P|> 2$ and $0\leq \t\leq 1/4$.
Explicit leading order expressions can be obtained for any  $0\leq \t\leq 1/4$.
They show that the portion of atoms in the
condensate with high axial momentum significantly increases (see
Fig. \ref{Benfig}) as the point of gradient catastrophe at $\t=0.25$
is approached.

\section{Calculation of semiclassical NLS solutions for general one-hump initial data}
\label{sec:calc-semicl-nls}

Equation \eqref{FNLS}, the integrable NLS, can be solved by the
inverse scattering technique.  However, the semiclassical limit
solutions require the {\it semiclassical limit} of the scattering
transform.  Let $\S$ be the curve in the upper half plane, defined
parametrically by the analytic initial data \eqref{IDe} as $\a(x)=-\hf
S'(x) +iA(x)$, $x\in\R$.  Here $A(x)$ and $S'(x)-\m_\pm$, where
$\m_\pm$ are some real numbers, have sufficient decay as
$x\ra\pm\infty$ respectively.  Let $z$ be a point on $\S$.  Assuming
for simplicity that $\a(x)$ is invertible, the semiclassical
scattering data limit $f_0(z)$, $z\in\S$, is defined \cite{TV3}
through the generalized Abel integral transform as
\begin{align}\label{xtofpart}
 f_0(z)=
\int^{\m_+}_{z}&\left[z-\m_+ + \sqrt{(z-u)(z-\bar u)} \right]x'(u)du\cr
 &+(z-\m_+) x(z),
\end{align}
where $x(\a)$ is inverse to $\a(x)$ and the integral is taken along
$\S$. The analytic extension of $f_0(z)$ from $\S$ to $\R$ (which can
have logarithmic branch cuts) has a meaning of the leading order term
of $\hf i\e\ln r_0(z,\e)$ as $\e\ra 0$, where $r_0(z,\e)$, $z\in\R$,
is the reflection coefficient of \eqref{IDe}.  Once $f_0(z)$ is known,
the complex wave parameters encoded in the branch points of
$\Rscr(x,\t)$ are defined through the modulation equations. In
particular, in the genus zero region, the modulation equation for
$\a(x,\t)\in\C^+$ is given by the system of two real equations
\cite{TVZ1}
\begin{equation}
\label{modeqg0}
\int_\gamma \frac{f'(\zeta)}{R (\zeta)} d\zeta =0,~~~~~~~
\int_\gamma \frac{\z f'(\zeta)}{R(\zeta)} d\zeta =0,
\end{equation}
where $f(z)=f(z;x,\t)=f_0(z)-xz-\t z^2$, $R(z)=\sqrt{(z-\a)(z-\bar
  \a)}$ and $\g$ is an oriented, Schwarz-symmetrical contour
connecting $\bar\a$ and $\a$, such that $\g \cup \R = \m_+$. It
defines $q(x,\t,\e)$ through \eqref{qg0}-\eqref{alph}.  Here $f_0(z)$
is Schwarz symmetrically extended into the lower half plane; typically
$\Im f_0(z)$ has a jump along $\R$.

We define the function $h(z)=h(z;x,\t)$ as
\be\label{h}
h(z)=\frac{R(z)}{i\pi}\int_{\g}{\frac{f(\z)}{(\z-z)R(\z)}}d\z-f(z).
\ee  
Because
of the analyticity of $f(z)$, the particular shape of 
{$\g$} is not important.  However, it is possible to fix $\g$ by
the condition $\Im h(z)=0$ on $\g$.  According to the Deift-Zhou
nonlinear steepest descent method \cite{DVZ}, the genus zero ansatz
\eqref{qg0} approximates the actual solution of the NLS \eqref{FNLS}
with the reflection coefficient $r_0(z,\e)=e^{-\frac{2i}{\e}f_0(z)}$
if (see \cite{TVZ1}) simultaneously:
\begin{align}\label{ineq}
\Im h(z;x,\t)&<0 ~{\rm on~ both ~sides~  of~ } \g^+;\cr
\Im h(z;x,\t)&>0 ~{\rm  on~ }\g^+_{c},
\end{align}
where $\g^+_c$ is a contour in the upper half plane $\C^+$ connecting
$\a$ and $\m_-$ and {$\g^+=\g \cap
  \C^+$}. We
have the freedom to deform the contours $\g,\g^+_c$ so that the
{inequalities}  \eqref{ineq} {are}
satisfied along it.  The first breaking curve
consists of points $(x,\t)$ where at least one of the inequalities
\eqref{ineq} turns into equality at some $z_0$. Thus, the equation for
the first breaking curve can be written as a system of three real
equations for $z_0\in\C$ and $(x,\t)\in\R^2$ 
\be\label{br1} \Im
h(z_0;x,\t)=0~~~~~~~~~~{\rm and}~~~~~~~~~~~~h_z(z_0;x,\t)=0~.
\ee

For the initial data (\ref{eq:2}) with $\mu = 2$ , the expression
\begin{equation}
  \label{hexpl}
  \begin{split}
    h(z)= &z\ln\frac{\sqrt{a^2+b^2}R(z)-a(z-a)+b^2}{z}- \\
    &\t(z-a) R(z)+(1-z)\left[\ln b-\ipt\right]-\\
    &\ln[R(z)-(z-a)],
  \end{split}
\end{equation}
was found in \cite{TVZ1}.  The modulation equations, as well as
expressions for $h(z;x,\t)$ for higher genus regions, can be written
in explicit determinantal form (see \cite{TV1,TV2}).  Since these
expressions are somewhat involved, they will not be given in this
paper.

The point of gradient catastrophe $x_0,\t_0$ is defined by the system of 
equations that consists of (\ref{modeqg0}) {and the
  following}  two equations (see \cite{TVZ3}): 
\begin{equation}
\label{modeqf''}
\int_\gamma \frac{f''(\zeta)}{R_+ (\zeta)} d\zeta =0,~~~~~~~
\int_\gamma \frac{\z f''(\zeta)}{R_+ (\zeta)} d\zeta =0.
\end{equation}
This is a system of four real equations for $\a\in\C$ and $(x_0,\t_0)\in\R^2$.

Equations (\ref{xtofpart})-(\ref{modeqf''}) show that the
$\mathcal{O}(\e)$-approximate solution in the genus zero region, the
point of gradient catastrophe, and the breaking curve can be
effectively calculated from decaying initial data of the form
(\ref{IDe}).  Further calculations also reveal the asymptotic
structure of the solution in the genus two region near the first
breaking curve and around the point of gradient catastrophe, see
\cite{BT1} and \cite{BT2} respectively.

\section{Suggestions and Conclusions}
\label{sec:sugg-concl}

The semiclassical limit of the {\it focusing} 1D NLS \eqref{FNLS}
provides a new,
mathematically rigorous tool to study the modulationally unstable
evolution of an attractive 1D BEC.  When the conditions
$N<N_\textrm{cr}$ (see eq. (\ref{eq:27})) and inequalities
(\ref{eq:15}), (\ref{eq:30})--or eq.~(\ref{eq:48})--are satisfied, an
attractive BEC in
an extended cylindrical trap (cigar shaped potential without axial
caps) is expected to be governed by the focusing NLS in the
semiclassical regime. A typical evolution for decaying, single-hump,
analytic initial data $q(x,0,\e)$, see (\ref{IDe}), is depicted in
Fig.~\ref{Cai}.
Using new tools from asymptotic inverse scattering theory, in
particular, the Deift-Zhou nonlinear steepest descent method, outlined
above, we were able to calculate the solution with $\mathcal{O}(\e)$
accuracy in several different regimes.
We have obtained a number of macroscopic characteristics of the
evolving condensate.
These include (see Fig. \ref{Cai}) the space-time location of the
point of gradient catastrophe and the breaking curves, the slowly
modulated amplitude in the genus zero region, and the envelope of the
fast amplitude oscillations in the genus two region.

Attractive 1D BEC experiments without axial trap are expected to
produce two counterpropagating radiative waves with the space in
between filled with two-phase modulated waves and, possibly, with
spatially localized coherent structures that consist of more
complicated nonlinear waves. These higher genus stationary structures
are expected to be linked with the discrete spectrum of the
corresponding ZS system.

The calculation of the observables, mentioned above, is based on the
semiclassical limit of the scattering data $f_0(z)$, which, in its
turn, can be obtained from $q(x,0,\e)$ through \eqref{xtofpart}, i.e.,
through the initial amplitude $A(x)$ and the phase $S(x)$. However,
accurate measurement of the initial phase is often a difficult task. We
can turn the question around and ask whether the phase $S(x)$ can be
somehow reconstructed from $A(x)$ and some observables. Continuation
of this line of argument leads to the question of designing some NLS
data, initial or scattering, whose evolution will have certain desired
properties and/or fit within some required parameters. In light of the
described above example of the initial amplitude $A(x)=\sech x$ and the
phase gradient $S'(x)=-\m\tanh x$, it seems to be especially
interesting to observe experimentally how increasing the phase
gradient $S'(x)$ for the same amplitude $A(x)$ (by, for example,
applying an external focusing potential) can reduce the number of
phase transitions of the evolving condensate.

Generally speaking, formulae \eqref{xtofpart}-\eqref{br1} are valid
for a large class of analytic initial data, including, for example,
multi-hump initial data.  Further theoretical and
  experimental investigations of multi-hump initial densities may
  yield interesting results.

The authors thank V. Kokoouline for stimulating discussions and
B. Relethford, who participated in the summer 2009 REU DMS 0649159 at the UCF
under the supervision of the first author, for
 calculating  the Fourier transform $\hat q(p,\t,\e)$  and creating
Figure \ref{Benfig}.

\appendix*
\section{Some energy estimates}
\label{sec:appendix}

In this appendix, we briefly outline the derivation of the quasi-1D
criterion (\ref{eq:15}).

The conserved energy associated with the 3D GP equation
(\ref{GPEfull}) for $t > 0$ (when $V(\rv,t) = V_\perp(y,z)$ and $g(t) =
4\pi \hbar^2 a_s^{(a)}/m < 0$) is
\begin{equation}
  \label{eq:5}
    \begin{split}
      \mathcal{E}[\Psi] = & ~ \mathcal{E}_\perp +
      \mathcal{E}_\parallel +
      \mathcal{E}_{\textrm{nl}} \\
      \equiv & ~ \int_{\R^3} \left \{ \frac{\hbar^2}{2 m} |
        \nabla_\perp' \Psi |^2
        + V_\perp(y',z') | \Psi |^2 \right \} d\rv' + \\
      & \int_{\R^3} \frac{\hbar^2}{2
        m} | \Psi_{\xi'} |^2 d \rv' + \int_{\R^3} \frac{g}{2} | \Psi
      |^4 d \rv' .
  \end{split}
\end{equation}
Assuming that $\Psi$ has the approximate separated form given in
eq.~(\ref{eq:29}), we formally compute
\begin{equation}
  \label{eq:6}
  \begin{split}
    \mathcal{E}_\perp &= \frac{N \omega_\perp \hbar}{2\pi} \| q(\cdot,
    \t)
    \|_{L^2(\R)}^2, \\
    \mathcal{E}_\parallel &= \frac{\hbar^2 N}{4
      \pi m \Delta^2} \| q_z(\cdot, \t) \|_{L^2(\R)}^2, \\
    \mathcal{E}_{\textrm{nl}} &= \frac{\hbar a_s^{(a)} N^2
      \omega_\perp}{2 \pi \Delta} \| q(\cdot, \t) \|_{L^4(\R)}^4 ,
  \end{split}
\end{equation}
where $\| f(\cdot) \|_{L^p(\R)} \equiv (\int_\R |f(x)|^p)^{1/p}$ is
the standard $L^p$ norm.  If the characteristic energies of the radial
harmonic oscillator excitations $\delta \mathcal{E}_\perp \ge
\mathcal{E}_\perp$ are much greater then the energies associated with
the axial $\mathcal{E}_\parallel$ and nonlinear
$\mathcal{E}_{\textrm{nl}}$ excitations, the 3D GP equation
\eqref{GPEfull} can be approximated by a 1D GP equation in the
longitudinal (axial) direction \cite{Abdul,Kevrekidis}.  For the
separated ansatz (\ref{eq:2}) and the 1D NLS equation (\ref{FNLS}) to
be valid, we therefore assume
\begin{equation}
  \label{eq:7}
  \mathcal{E}_\perp \gg \mathcal{E}_\parallel, \quad \mathcal{E}_\perp
  \gg |\mathcal{E}_{\textrm{nl}}| ,
\end{equation}
or, since the $L^2$ norm of $q$ is unity (see eq.~(\ref{eq:28})),
\begin{equation}
  \label{eq:32}
  \begin{split}
    \frac{a_0^2}{2 \Delta^2} \| q_z(\cdot,\t) \|_{L^2(\R)}^2 \ll 1,
    \quad \frac{|a_s^{(a)}| N}{\Delta} \| q(\cdot,\t)
    \|_{L^4(\R)}^4 \ll 1 .
  \end{split}
\end{equation}
Further simplification can be made by using the fact that the 1D
energy for $q$ satisfying the NLS equation (\ref{FNLS}) is conserved
\begin{equation}
  \label{eq:33}
  \mathcal{E}_{\textrm{1D}} = \frac{\e^2}{2} \| q_z(\cdot,\t) \|_{L^2(\R)}^2
  - \frac{1}{2} \| q(\cdot,\t) \|_{L^4(\R)}^4 .
\end{equation}
For slowly varying, single-hump (Gaussian type) initial data in the
semiclassical regime (\ref{eq:30}) satisfying (\ref{eq:28}), we will
have $\mathcal{E}_{\textrm{1D}} < 0$.  Then the inequalities in
(\ref{eq:7}) become
\begin{equation}
  \label{eq:22}
  \begin{split}
    \frac{N |a_s^{(a)}|}{\Delta} (\| q(\cdot,\t) \|_{L^4(\R)}^4 - 2 |
    \mathcal{E}_{\textrm{1D}}| ) &\ll 1, \\
    \frac{N |a_s^{(a)}|}{\Delta} \| q(\cdot,\t) \|_{L^4(\R)}^4 &\ll 1 .
  \end{split}
\end{equation}
If we have
\begin{equation}
  \label{eq:8}
  \max_{x \in \R} (|q(x,\t)|^2) \le C, \quad \t \in [0,T_0],
\end{equation}
then the $L^4$ norm of $q$ is bounded by $C$ because
\begin{equation}
  \label{eq:4}
  \| q(\cdot,\t) \|_{L^4(\R)}^4 \le \max_{x \in \R} (|q(x,\t)|^2)
  \|q(\cdot,\t) \|^2_{L^2(\R)} \le C , 
\end{equation}
for $\tau \in [0,T_0]$ by use of eq.~(\ref{eq:28}).  Then the
inequalities in (\ref{eq:22}) leading to a quasi-1D BEC are satisfied
when eq.~(\ref{eq:15}) holds. Using \eqref{eq:28} and our
semiclassical calculations, we can estimate $C \approx
9|q(x_0,\t_0)|^2$ for $\t \in [0,T_0]$ with $T_0>\t_0$ but $T_0 <
\t_1$, the time of the second break.  For example, the initial data in
eq.~(\ref{eq:2}) give $C \lesssim 9(\mu + 2)$.

\end{document}